\documentclass[epj]{webofc}
\usepackage[utf8]{inputenc}
\usepackage[varg]{txfonts}   % Web of Conferences font
\usepackage{booktabs}
\usepackage{xcolor}
\definecolor{darkred}{rgb}{0.4,0.0,0.0}
\definecolor{darkgreen}{rgb}{0.0,0.4,0.0}
\definecolor{darkblue}{rgb}{0.0,0.0,0.4}
\usepackage[bookmarks,linktocpage,colorlinks,
    linkcolor = darkred,
    urlcolor  = darkblue,
    citecolor = darkgreen]{hyperref}

\wocname{EPJ Web of Conferences}
\woctitle{Lattice2017}

\newcommand{\sixth}{\mbox{\small $\frac{1}{6}$}}         % 1/6
\newcommand{\half}{\mbox{\small $\frac{1}{2}$}}          % 1/2
\newcommand{\third}{\mbox{\small $\frac{1}{3}$}}         % 1/3
\newcommand{\twothird}{\mbox{\small $\frac{2}{3}$}}      % 2/3
\newcommand{\oosqrttwo}{\mbox{\small $\frac{1}{\sqrt{2}}$}}% 1/sqrt(2)
\newcommand{\oosqrtsix}{\mbox{\small $\frac{1}{\sqrt{6}}$}}% 1/sqrt(6)

\colorlet{alert}{red!60!black}
\colorlet{example}{green!60!black}
\colorlet{structure}{blue!60!black}

\begin{document}

%%%%%%%%%%%%%%%%%%%%%%%%%%%%%%%%%%%%%%%%%%%%%%%%%%%%%%%%%%%%%%%%%%%%%%%%%%%%%

%
\selectlanguage{english}

%----------------------------------------------------------------------------

\title{Charmed states and flavour symmetry breaking}

%\subtitle{-- QCDSF/UKQCD Collaborations --}

%----------------------------------------------------------------------------

\author{
\lastname{R.~Horsley}\inst{1}\fnsep\thanks{Speaker, \email{rhorsley@ph.ed.ac.uk}}
\and
\lastname{Z.~Koumi}\inst{2}
\and
\lastname{Y.~Nakamura}\inst{3}
\and
\lastname{H.~Perlt}\inst{4}
\and
\lastname{P.~E.~L.~Rakow}\inst{5}
\and
\lastname{G.~Schierholz}\inst{6}
\and
\lastname{A.~Schiller}\inst{4}
\and
\lastname{H.~St\"uben}\inst{7}
\and
\lastname{R.~D.~Young}\inst{2}
\and
\lastname{J.~M.~Zanotti}\inst{2}
\\ for the QCDSF-UKQCD Collaborations
%
%\firstname{Bilbo} \lastname{Baggins}\inst{1,3}\fnsep\thanks{Acknowledges financial support by his mentor J.R.R. Tolkien.} \and
%\firstname{Harry} \lastname{Potter}\inst{2} \and
%\firstname{Star}  \lastname{Lord}\inst{3}\fnsep\thanks{Speaker, \email{guardians@galaxy.net} (only for submitting author)}
%
}
%
% DESY 17-181 
% Edinburgh 2017/23
% Adelaide preprint number: ADP-17-35/T1041
% Liverpool preprint number: LTH 1144 
%
%----------------------------------------------------------------------------

\institute{
School of Physics and Astronomy,
University of Edinburgh, Edinburgh EH9 3FD, UK
\and
CSSM, Department of Physics,
University of Adelaide, Adelaide SA 5005, Australia
\and
RIKEN Advanced Institute for Computational Science,
Kobe, Hyogo 650-0047, Japan
\and
Institut f\"ur Theoretische Physik,
Universit\"at Leipzig, 04109 Leipzig, Germany
\and
Theoretical Physics Division, Department of Mathematical Sciences,
University of Liverpool, Liverpool L69 3BX, UK
\and
Deutsches Elektronen-Synchrotron DESY,
22603 Hamburg, Germany
\and
Regionales Rechenzentrum, Universit\"at Hamburg,
20146 Hamburg, Germany
}

%----------------------------------------------------------------------------

\abstract{
   Extending the SU(3) flavour symmetry breaking expansion
   from up, down and strange sea quark masses to partially quenched
   valence quark masses allows an extrapolation to the charm quark
   mass. This approach leads to a determination of charmed quark 
   hadron masses and decay constants. We describe our recent progress
   and give preliminary results in particular with regard to the
   recently discovered doubly charmed baryon by the 
%   recently discovered doubly charmed baryon (the $\Xi_{cc}^{++}$) by the 
%   recently discovered doubly charmed baryon (the Xicc++) by the 
   LHCb Collaboration.
}

\hfill ADP-17-35/T1041 \,\,\, DESY 17-181 \,\,\,  
       Edinburgh 2017/23 \,\,\, Liverpool LTH 1144

% DESY 17-181 
% Edinburgh 2017/23
% Adelaide preprint number: ADP-17-35/T1041
% Liverpool preprint number: LTH 1144 

%----------------------------------------------------------------------------

\maketitle

%----------------------------------------------------------------------------

\section{Introduction}
\label{intro}

%----------------------------------------------------------------------------

Open charm baryon masses are presently the subject of much experimental
and theoretical interest. In particular, most doubly charmed baryons
(i.e.\ $ccq$ states) have not been experimentally observed even though
as stable states in QCD they must exist. Indeed there has only been
the observation of a candidate state, $\Xi_{cc}^+$ (with quark content
$ccd$) by the SELEX Collaboration, \cite{selex02a}. However this state
was not seen by the BaBar, \cite{aubert06a} or BELLE, \cite{chistov06a}.
Collaborations. Very recently, however, the LHCb Collaboration,
\cite{LHCb17a}, has announced a state -- the $\Xi_{cc}^{++}$ baryon -- 
with a quark content $ccu$. It is unlikely that isospin breaking effects
are significant (i.e.\ QED effects and $m_u \not= m_d$), 
or that the states have been missidentified
so between the SELEX and LHCb result is an unexplained and puzzling
mass difference of $\sim 100\,\mbox{MeV}$.

In this talk we shall describe the QCDSF-UKQCD approach to determining the
hadron mass spectrum, with particular emphasis on the charm sector
and the open doubly charmed baryon masses. This continues the programme
initialised in \cite{horsley13a}.

$2+1$ flavour dynamical lattice simulations consist of two mass
degenerate (i.e.\ $m_u = m_d$) light flavour $u$, $d$ quarks and a
heavier flavour $s$ quark. Since the light quark masses are typically 
larger than the `physical' masses required for the experimental spectrum,
we are forced to consider how we can approach the physical $u$, $d$, $s$
quark masses. While many current simulations determine the physical
$s$ quark mass and then extrapolate the $u$, $d$ quark masses to the
physical quark mass, another possibility as suggested in \cite{bietenholz11a}
is to consider an SU(3) flavour breaking expansion from a point
$m_0$ on the flavour symmetric line keeping the average quark mass
$\overline{m} = (m_u+m_d+m_s)/3$ constant.
This procedure significantly reduces the number of expansion
coefficients allowed. (Also, not considered here, the expansion
coefficients remain the same whether we consider $m_u \not= m_d$
or $m_u = m_d$ and thus allows for the possibility of finding
the pure QCD contribution to isospin breaking effects using just
$n_f = 2+1$ numerical simulations.)

As the charm quark is considerably heavier than the up, down and
strange quarks an SU(4) flavour breaking expansion is poorly
convergent (in distinction to the SU(3) flavour breaking expansion).
Another possibility is to make independent SU(3) flavour breaking expansions
in each charm sector (but this not a very unified approach).
We adopt an intermediate aproach here, first noting that as the
charm quark mass is much heavier than the $u$, $d$ and $s$ quark masses,
it contributes little to the dynamics of the sea of the hadron.
Thus we can regard the charm quark as a `Partially Quenched' or PQ quark.
By this we mean that the quark masses making up the meson or baryon
have not necessarily the same mass as the sea quarks.
Also the SU(3) flavour breaking expansion can also be extended
to valence quark masses. As the expansion coefficients are just
functions of $\overline{m}$, provided this is kept constant then
the coefficients are unchanged. We shall say the `Unitary Limit' when
the masses of the valence quarks coincide with the sea quarks.
PQ determinations have the advantage of not being
expensive compared to dynamical simulations of the quarks. This can
also help in the determination of the expansion coefficients as
a wider range of quark masses than just the unitary masses can be used.

Briefly the method employed here is to first determine the expansion
coefficients of the pseudoscalar mesons and octet baryons.
Extrapolating to the physical pseudoscalar masses determines the
`physical' quark masses, which are then employed in the baryon
expansions to determine the (open) charm masses.

%----------------------------------------------------------------------------

\section{SU(3) flavour breaking expansions}

%----------------------------------------------------------------------------

We shall only consider here hadrons which lie on the outer ring
of their associated multiplet and not the central hadrons. So no
mixing or quark--line disconnected correlation functions are considered
here. The SU(3) flavour symmetry breaking expansions are given
in terms of
\begin{eqnarray}
   \delta \mu_q  = \mu_q - \overline{m} \,, \quad
            \overline{m} = \third(2m_l + m_s) \,,
                                                           \nonumber 
\end{eqnarray}
where $\mu_q$ are the PQ masses for quark $q$. In the unitary limit
we have $\mu_q \to m_q$, where $m_q$ is the unitary quark mass (i.e.\ equal
to the sea quark mass). Here we have the obvious constraint
$2\delta m_l + \delta m_s = 0$.

For the pseudoscalar mesons with valence quarks $q = a$ and $b$, their
masses are given to NLO (or quadratic in $\delta\mu_q$ by
\begin{eqnarray}
   M^2(a\overline{b})
      = M^2_{0\pi} + \alpha(\delta\mu_a + \delta\mu_b)
            + \beta_0\sixth(2\delta m_l^2 + \delta m_s^2)
            + \beta_1(\delta\mu_a^2 + \delta\mu_b^2)
            + \beta_2(\delta\mu_a - \delta\mu_b)^2 + \ldots \,,
                                                         \nonumber
\label{M2ps_expan}
\end{eqnarray}
(cubic or NNLO terms are given in \cite{horsley12a}). The expansion
coefficients are functions of $\overline{m}$ only, so if we keep
$\overline{m}$ fixed we have constrained fits between the pseudoscalar
mesons. 

Numerically it is advantageous to use scale invariant quantities.
A useful additional quantity with this method is to use flavour singlet
or blind quantities, only defined in the unitary limit. There are many
possibilities, for example for pseudoscalar meson quantities a 
convenient one is
\begin{eqnarray}
   X_\pi^2 = \third(2M_K^2 + M_\pi^2)
          = M_{0\pi}^2 + O(\delta m_l^2) \,,
                                                           \nonumber
\end{eqnarray}
Now forming dimensionless ratios, $\tilde{M}^2 = M^2/X_\pi^2$,
$\tilde{\alpha} = \alpha/M_{0\pi}^2$, 
$\tilde{\beta}_i = \beta_i/M_{0\pi}^2 \,, \ldots$ leads to the modified
expansion
\begin{eqnarray}
   \tilde{M}^2(a\overline{b})
      = 1 + \tilde{\alpha}(\delta\mu_a + \delta\mu_b)
            - (\twothird\tilde{\beta}_1 + \tilde{\beta}_2)
                   (2\delta m_l^2 + \delta m_s^2)
           + \tilde{\beta}_1(\delta\mu_a^2 + \delta\mu_b^2)
           + \tilde{\beta}_2(\delta\mu_a - \delta\mu_b)^2 + \ldots \,.
                                                         \nonumber
\label{Mps2twid_expan}
\end{eqnarray}
Similarly for the outer ring of the baryon octet, we have the expansion
\begin{eqnarray}
   \tilde{M}_{\Sigma}^2(aab)
      &=& 1 + \tilde{A}_1(2\delta\mu_a + \delta\mu_b) 
              + \tilde{A}_2(\delta\mu_b - \delta\mu_a)
                                                         \nonumber   \\
      & &     - (\tilde{B}_1 + \tilde{B}_3) (2\delta m_l^2+\delta m_s^2)
              + \tilde{B}_1(2\delta\mu_a^2 + \delta\mu_b^2)
              + \tilde{B}_2(\delta\mu_b^2 - \delta\mu_a^2) 
              + \tilde{B}_3(\delta\mu_b - \delta\mu_a)^2 \,,
% + \ldots \,,
                                                         \nonumber
\label{M2N_expan}
\end{eqnarray}
where we have collectively denoted these baryons with a $\Sigma$ index.
Similarly to the meson case, we have formed dimensionless quantities
by normalising with the singlet quantity
\begin{eqnarray}
   X_N^2 = \third( M_N^2 + M_\Sigma^2 + M_\Xi^2 ) 
         = M_{0N}^2 + O(\delta m_l^2) \,.
                                                         \nonumber
\label{XN_def}
\end{eqnarray}
The expansion given here to NLO (quadratic in $\delta\mu_q$).
The expansion to NNLO (at next order with is given, for example,
in \cite{horsley14a} Appendix C. NNLO is used in the analysis presented here.
Provided $m_u = m_d$ then for the $\Lambda$ we have
\begin{eqnarray}
   \tilde{M}^2_{\Lambda}(aa^\prime b)
      &=& 1 + \tilde{A}_1(2\delta\mu_a + \delta\mu_b) 
            - \tilde{A}_2(\delta\mu_b - \delta\mu_a)
                                                         \nonumber   \\
      & &   - (\tilde{B}_1 + \tilde{B}_3) (2\delta m_l^2+\delta m_s^2)
            + \tilde{B}_1(2\delta\mu_a^2 + \delta\mu_b^2)
            - \tilde{B}_2(\delta\mu_b^2 - \delta\mu_a^2) 
            + \tilde{B}_4(\delta\mu_b - \delta\mu_a)^2 \,,
%+ \ldots
                                                         \nonumber
\label{M2L_expan}
\end{eqnarray}
(the NNLO terms are also given in \cite{horsley14a}. We shall use a prime,
such as $a^\prime$, to denote a distinct quark, but with the same mass as 
quark $a$. (It turns out that the fits are slightly better if the square
of the mass is used rather than just the mass. The SU3 flavour breaking
expansions are valid for any function of the mass.)

Finally note that for open charm masses investigating mass splittings
can give information on SU(3) mass splittings, as to LO there is no
influence from the charm quark mass. For example for the pseudoscalar
mesons
\begin{eqnarray}
   \tilde{M}(a\overline{c}) - \tilde{M}(b\overline{c}) 
      = \half \tilde{\alpha}(\delta\mu_a + \delta\mu_b) + \ldots \,,
                                                         \nonumber
\end{eqnarray}
while for the octet baryons we have
\begin{eqnarray}
   \tilde{M}(aac) - \tilde{M}(bbc)
      &=& (\tilde{A}_1 - \half\tilde{A}_2) (\delta\mu_a-\delta\mu_b)
                   + \ldots \,,
                                                         \nonumber  \\
   \tilde{M}(cca) - \tilde{M}(ccb)
      &=& \half(\tilde{A}_1+\tilde{A}_2)(\delta\mu_a-\delta\mu_b) 
                + \ldots \,.
                                                         \nonumber
\end{eqnarray}

%----------------------------------------------------------------------------

\section{Lattice}

%----------------------------------------------------------------------------

We use a $O(a)$ non-perturbatively improved clover fermion action, together
with a Symanzik tree level improved glue. The fermions are also mildly
stout smeared. Further details may be found in \cite{cundy09a}. Thus
the quark mass is given by
\begin{eqnarray}
   \mu_q = {1 \over 2}
           \left( {1\over \kappa_q} - {1\over \kappa_{0c}} \right) \,,
                                                             \nonumber
\end{eqnarray}
where $\kappa_q$ is the hopping parameter, $\kappa_0$ is the
hopping parameter along the symmetric line with $\kappa_{0c}$ being
its chiral limit. Note that for $\delta\mu_q$, the `distance' from
the initial SU3 flavour symmetry point, $\kappa_{0c}$, cancels
and so does not have to be determined. We have presently analysed four
lattice spacings where $a \sim 0.052$ --  $0.074\,\mbox{fm}$,
but are aiming for five spacings.

\begin{figure}[!h]
\begin{center}

\begin{minipage}{0.45\textwidth}

   \vspace*{-0.25in}

   \begin{center}
      \includegraphics[width=6.25cm]
          {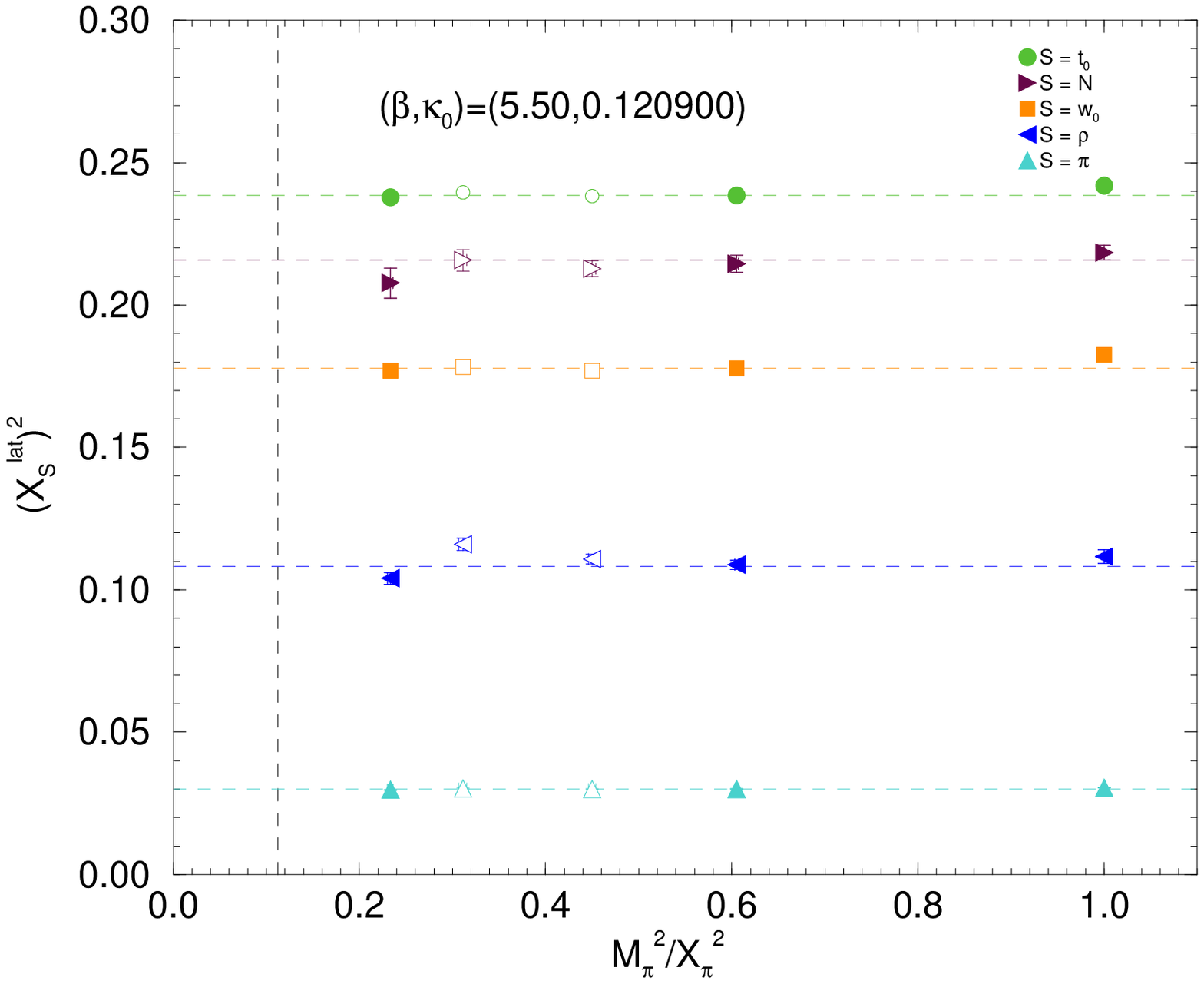}
   \end{center} 

\end{minipage}\hspace*{0.05\textwidth}
\begin{minipage}{0.45\textwidth}

   \begin{center}
      \includegraphics[width=6.25cm]{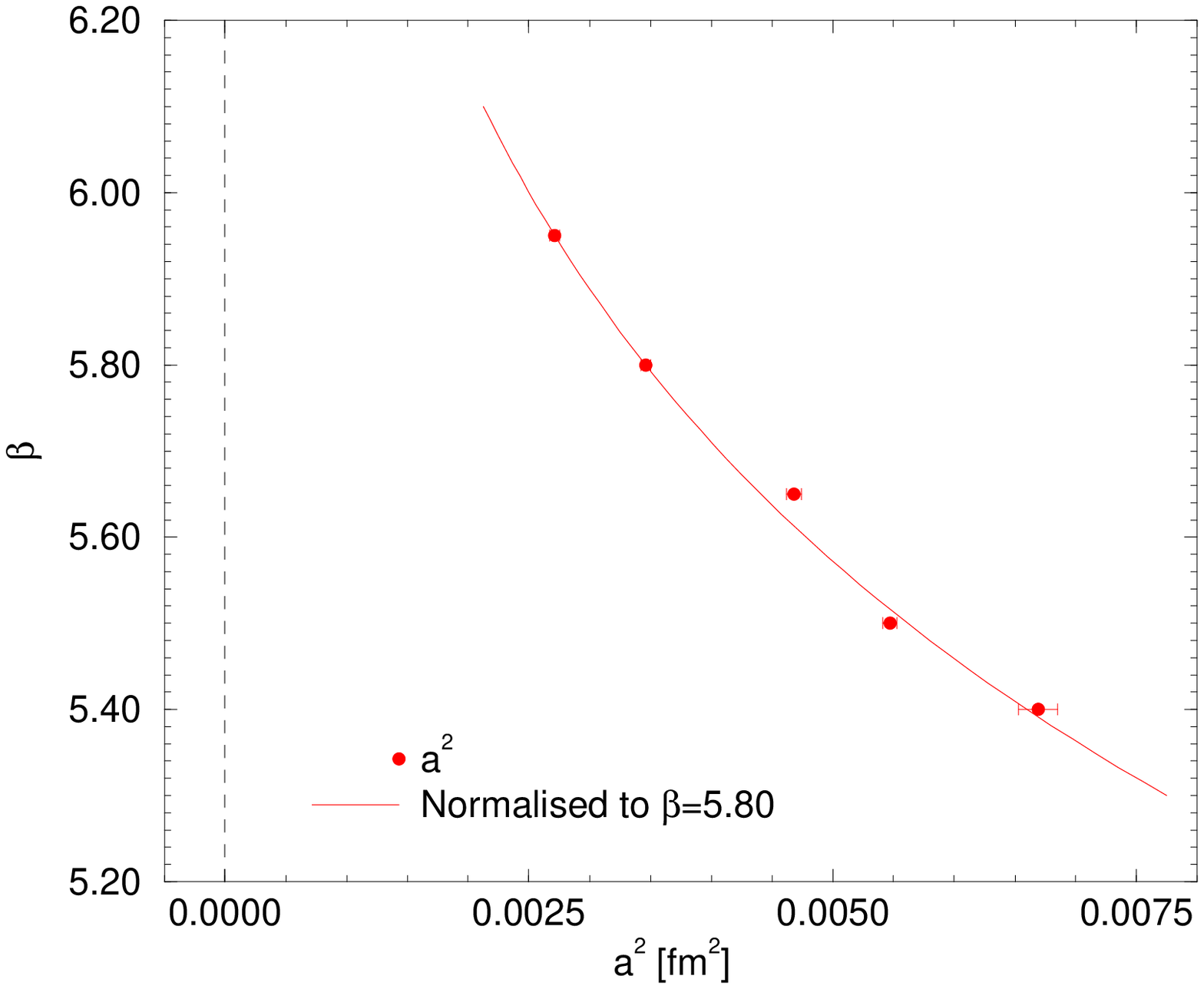}
   \end{center} 

\end{minipage}

\end{center}
\caption{Left panel: $X^{\rm lat}_S)^2$ for $S = t_0$, $N$, $w_0$,
         $\rho$ and $\pi$ along the unitary line,
         from the symmetric point $\delta m_l = 0$ down
         to the physical point
         (vertical dashed line) together with constant fits
         (for $\beta = 5.50$, $\kappa_0 = 0.120900$).
         Right panel: $a^2(\beta)$ values normalised to $a^2(5.80)$
         ($a^2 \sim 0.035\,\mbox{fm}^2$). The line shows the two-loop
         $\beta$-function.}
\label{X2_const}

\end{figure}

In the LH panel of Fig.~\ref{X2_const} we show $X_S^{{\rm lat}\,2}$ for various
singlet quantities, $S = t_0$, $N$, $w_0$, $\rho$, $\pi$. This can be used
to determine both the lattice spacing and $\kappa_0$ -- the point on the
SU(3) flavour symmetric, to start the path to the physical point.
\cite{bornyakov15a}. For example, this could be achieved by tuning 
$X_\pi^{{\rm lat}\,2}/X_N^{{\rm lat}\,2}$ to its `physical' value, using 
wherever possible
the `pure' QCD values in FLAG3, \cite{Aoki:2016frl}, otherwise from the PDG,
\cite{PDG17a}. In the RH panel we show $a^2$ values normalised by the value
at $\beta = 5.80$ using the two loop $\beta$-function.

\begin{figure}[h]
\begin{center}

\begin{minipage}{0.45\textwidth}

   \begin{center}
      \includegraphics[width=6.25cm]
          {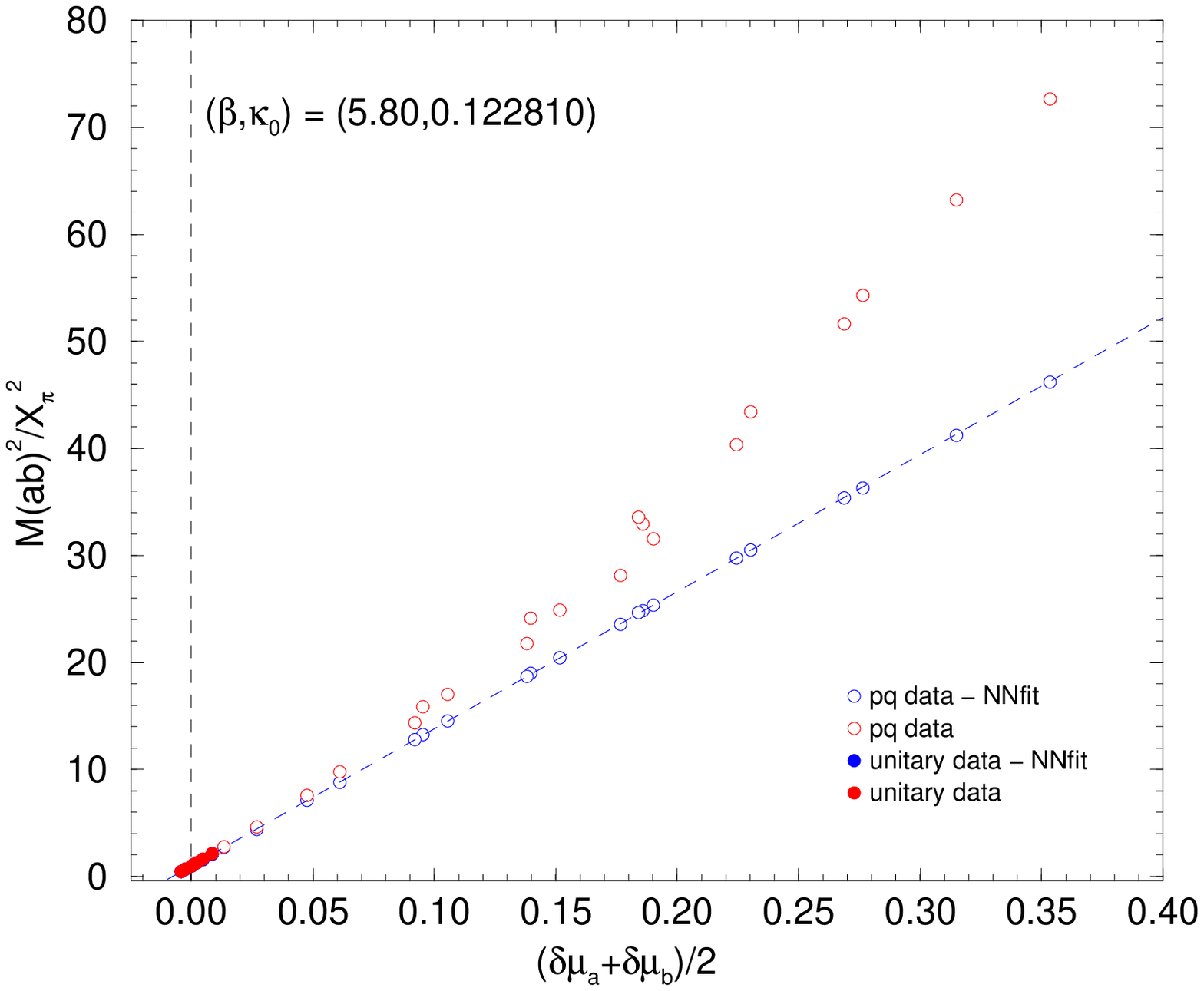}
   \end{center} 

\end{minipage}\hspace*{0.05\textwidth}
\begin{minipage}{0.45\textwidth}

   \begin{center}
      \includegraphics[width=6.25cm]
          {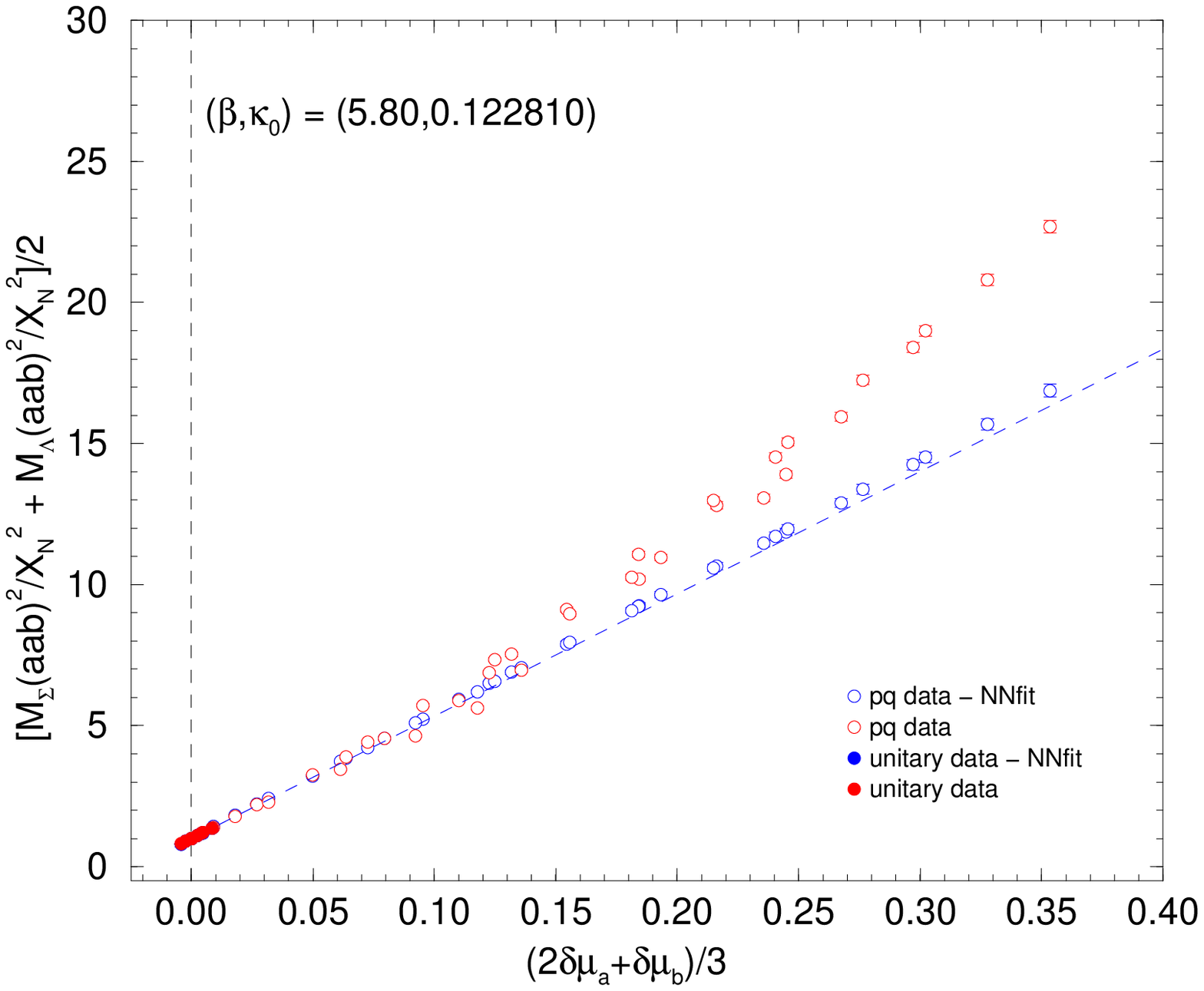}
   \end{center} 

\end{minipage}
\end{center}

\caption{Left panel: $\tilde{M}(a\overline{b}) = M^2(a\overline{b})^2/X_\pi^2$
         versus $(\delta\mu_a+\delta \mu_b)/2$ for 
         $(\beta,\kappa_0) = (5.80, 0.122810)$. The PQ masses are given
         by opaque red circles (the unitary data by filled red circles).
         Removing the NN term (leaving the LO terms) gives the blue
         opaque circles and linear fit term.
         Right panel: Similarly for 
         $(\tilde{M}_{\Sigma}^2(aab) + \tilde{M}_\Lambda^2(aa^\prime b))/2$
         versus $(2\delta\mu_a+\delta\mu_b)/3$.}
\label{expan_coeff}

\end{figure}

We first determine the pseudoscalar meson expansion coefficients.
In the LH panel of Fig.~\ref{expan_coeff} we show
$\tilde{M}(a\overline{b}) = M^2(a\overline{b})^2/X_\pi^2$
versus $(\delta\mu_a+\delta_b)/2$ (for $(\beta,\kappa_0) = (5.80, 0.122810)$).
To illustrate the fit (which at LO is a function $(\delta\mu_a+\delta \mu_b)/2$
only) after fitting we subtract the higher order terms to leave only
the LO term. Similarly in the RH panel of Fig.~\ref{expan_coeff} we
show the equivalent result for the octet baryon case by plotting
$(\tilde{M}_{\Sigma}^2(aab) + \tilde{M}_\Lambda^2(aa^\prime b))/2$
versus $(2\delta\mu_a+\delta\mu_b)/3$. To determine $\delta m_l^*$
the `physical' quark mass we set the pseudoscalar meson combination
$M_\pi^2/X_\pi^2$ to its physical value. Similarly we use 
$M_{\eta_c}^2/X_\pi^2$ to determine $\delta m_c^*$. 

Although we have scanned in detail the appropriate region
to determine the initial point, $\kappa_0$ for the trajectory some fine
tuning is possible. Bearing in mind that changes will come mostly
from the valence quark mass (rather than the sea quark masses), we
do not consider exactly the unitary limit, but allow $\delta\mu_l^*$,
$\delta\mu_s^*$ to be slightly different to $\delta m_l^*$, $\delta m_s^*$
by fitting $X_\pi^{pq\,2}/X^{pq\,2}_N$ and $M_\pi^{pq\,2}/X^{pq\,2}_\pi$ to their
physical values, while keeping the splitting between $\delta \mu_l^*$,
$\delta \mu_s^*$ the same as $\delta m_l^*$, $\delta m_s^*$. Presently
we see little difference, \cite{QCDSF18a}, and shall not discuss this
further here.

%----------------------------------------------------------------------------

\section{Results}

%----------------------------------------------------------------------------

\begin{figure}[!h]
\begin{center}

\begin{minipage}{0.45\textwidth}

   \begin{center}
      \includegraphics[width=6.25cm]
          {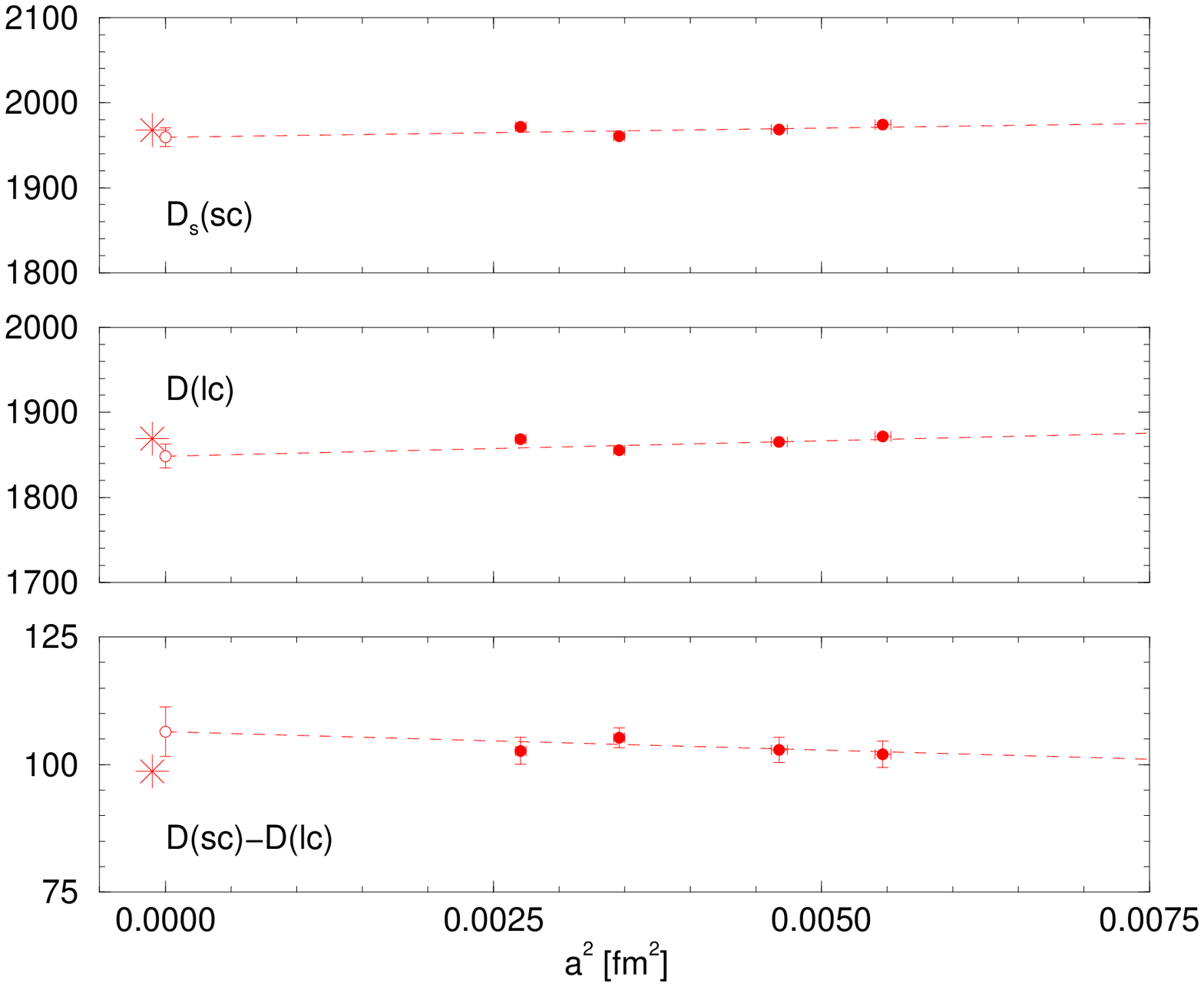}
   \end{center} 

\end{minipage}\hspace*{0.05\textwidth}
\begin{minipage}{0.45\textwidth}

   \begin{center}
      \includegraphics[width=6.25cm]{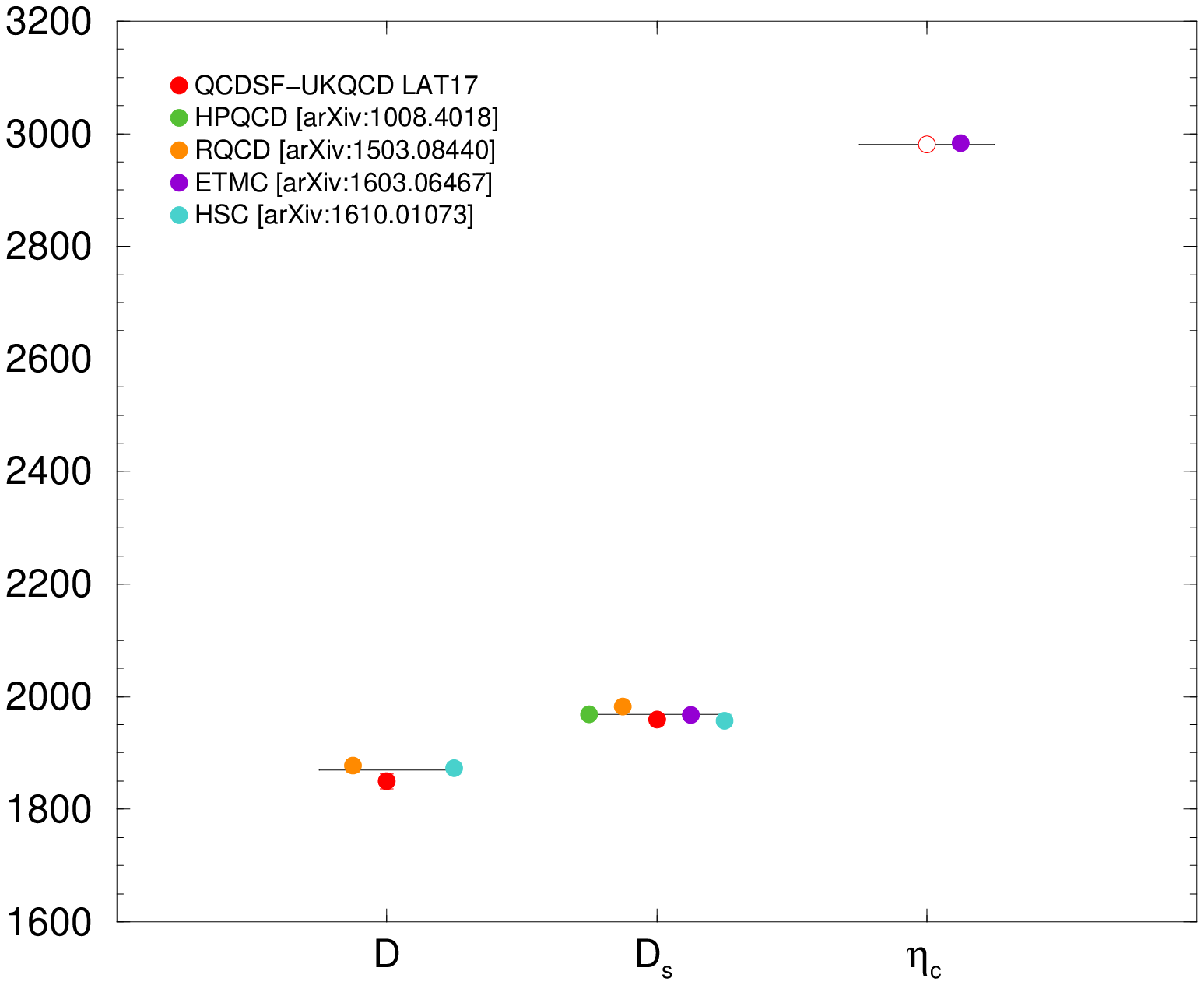}
   \end{center} 

\end{minipage}

\end{center}
\caption{Left panel: Preliminary $D_s(s\overline{c})$,
         and $D(l\overline{c})$ masses (upper and middle plot)
         together with their mass difference (lower plot)
         versus $a^2$, together with linear fits in $a^2$.
         The experimental results are shown with a star
         (slightly displaced in the -ve--$x$ direction for clarity).
         Right panel: Comparison of the extrapolated results 
         to other recent results, 
         \protect\cite{Davies:2010ip,Bali:2015lka,Cichy:2016bci,Cheung:2016bym}.
         The opaque red circle for the $\eta_c$ mass is just to
         indicate that this has been used in the determination of
         the charm quark mass.}

\label{D0_DD_pD_ps}

\end{figure}

After having determined the expansion coefficients and the `physical'
quark masses, we can first determine the open charm pseudsoscalar mesons.
All errors shown in the following plots are statistical
(however we expect systematic errors, if any, to be small).
In the LH panel of Fig.~\ref{D0_DD_pD_ps} we show preliminary results for
the continuum extrapolation of the $D_s(s\overline{l})$, 
$D(c\overline{l})$ masses and their mass difference
(upper to lower plots). Also shown is a linear extrapolation
in $a^2$ to the continuum limit. The phenomenological values
are denoted by a star. The mass differences in particular
are sensitive to unknown $u-d$ quark mass difference and 
QED effects (the present computation is for pure QCD only).
From the plots we see that there do not seem to be strong
scaling violations present. The RH panel of Fig.~\ref{D0_DD_pD_ps}
shows a comparison with some other determinations.

Presently we have determined charm baryon states with nucleon-like
wavefunctions ${\cal B} = \epsilon q(q^TC\gamma_5c)$ ($q = l$, $s$).
In Table~\ref{charm_wfs} we show the possible states in the isospin
\begin{table}[!htb]
   \begin{center}
      \begin{tabular}{ccrcc}
         $C$  &  $S$  &  $I$     &  baryon   &  wavefunction  \\
         \hline
          0   &  0    &  $\half$ &  $N(lll^\prime)$ & $\epsilon(l^TC\gamma_5l^\prime)l$  \\
          0   &  1    &  $1$     & $\Sigma(lls)$   & $\epsilon(l^TC\gamma_5s)l$  \\
          0   &  2    &  $\half$ & $\Xi(ssl)$      & $\epsilon(s^TC\gamma_5l)s$  \\
          0   &  1    &  $0$     & $\Lambda(ll^\prime s)$ & $\oosqrtsix\epsilon[2(l^TC\gamma_5l^\prime)s 
                                                           + (l^TC\gamma_5s)l^\prime - (l^{\prime\,T}C\gamma_5s)l]$  \\
         \hline
          1   &  0    &  $1$     & $\Sigma_c(llc)$ & $\epsilon(l^TC\gamma_5c)l$  \\
          1   &  1    &  $\half$ & $\Xi_c^{\prime}(lsc)$ & $\oosqrttwo\epsilon[(s^TC\gamma_5c)l + (l^TC\gamma_5c)s]$  \\
          1   &  2    &  $0$     & $\Omega_c(ssc)$ & $\epsilon(s^TC\gamma_5c)s$  \\
          1   &  0    &  $0$     & $\Lambda_c(ll^\prime c)$& $\oosqrtsix\epsilon[2(l^TC\gamma_5l^\prime)c 
                                                           + (l^TC\gamma_5c)l^\prime - (l^{\prime\,T}C\gamma_5c)l]$  \\
          1   &  1    &  $\half$ & $\Xi_c(csl)$    & $\oosqrtsix\epsilon[2(s^TC\gamma_5l)c 
                                                     + (s^TC\gamma_5c)l - (l^TC\gamma_5c)s]$  \\
         \hline
          2   &  0    &  $\half$ & $\Xi_{cc}(ccl)$ & $\epsilon(c^TC\gamma_5l)c$  \\
          2   &  1    &  $0$     & $\Omega_{cc}(ccs)$& $\epsilon(c^TC\gamma_5s)c$  \\
      \end{tabular}
   \end{center}
\caption{The possible $C=0$, $1$ and $2$ baryon octet states in
         the isospin symmetric limit, $m_u = m_d \equiv m_l$. A prime is
         used to denote a distinct quark, but with the same mass.}
\label{charm_wfs}
\end{table}
symmetric limit, $m_u = m_d \equiv m_l$. (A prime denotes a distinct
quark in the wavefunction, but with the same mass.) Thus in the charm
sector, we have presently investigated: $\Sigma_c(llc)$, $\Omega_c(ssc)$,
$\Xi_{cc}(ccl)$ and $\Omega_{cc}(ccs)$. 

\begin{figure}[h]
\begin{center}

\begin{minipage}{0.45\textwidth}

   \begin{center}
      \includegraphics[width=6.25cm]
          {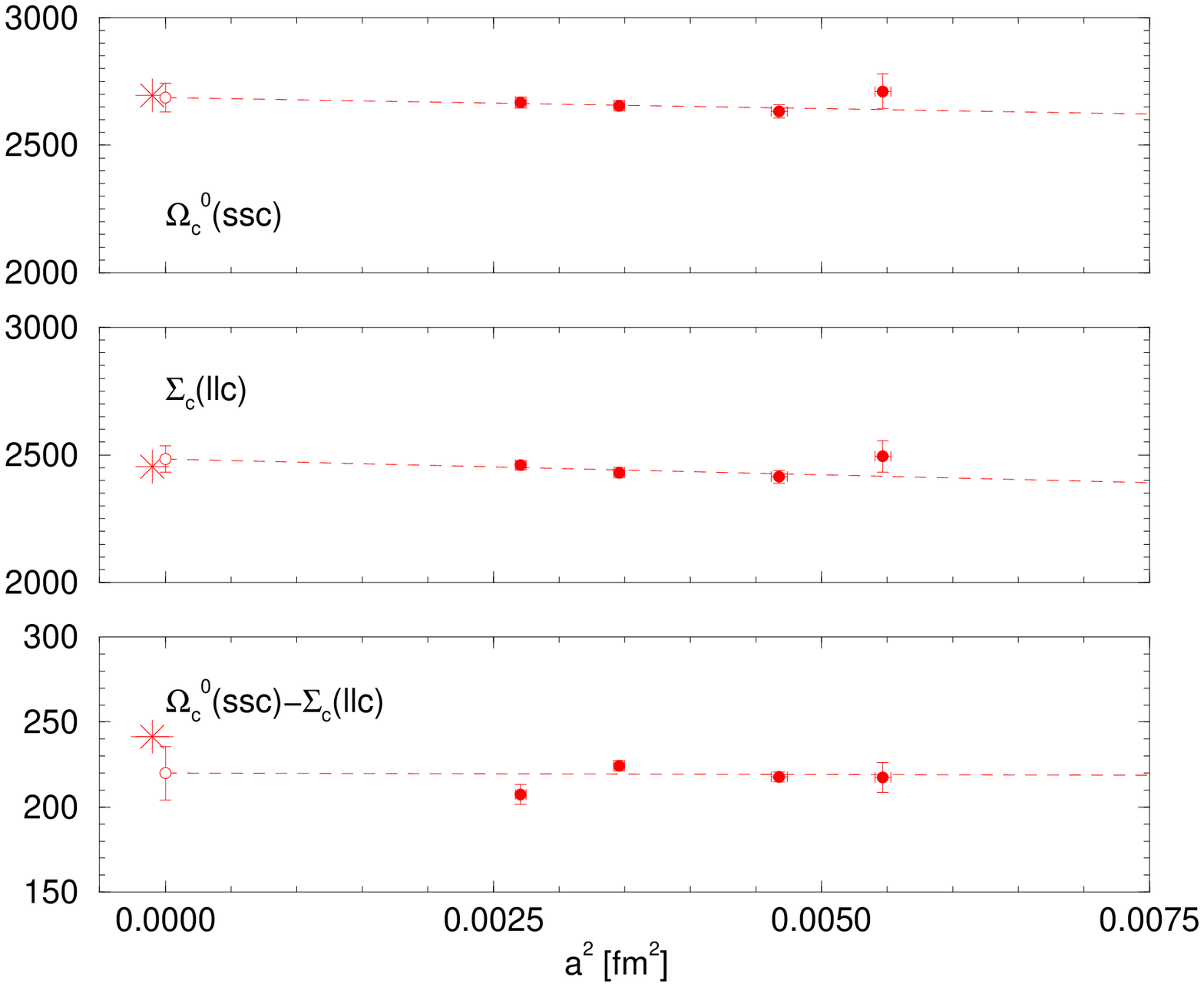}
   \end{center} 

\end{minipage}\hspace*{0.05\textwidth}
\begin{minipage}{0.45\textwidth}

   \begin{center}
      \includegraphics[width=6.25cm]
         {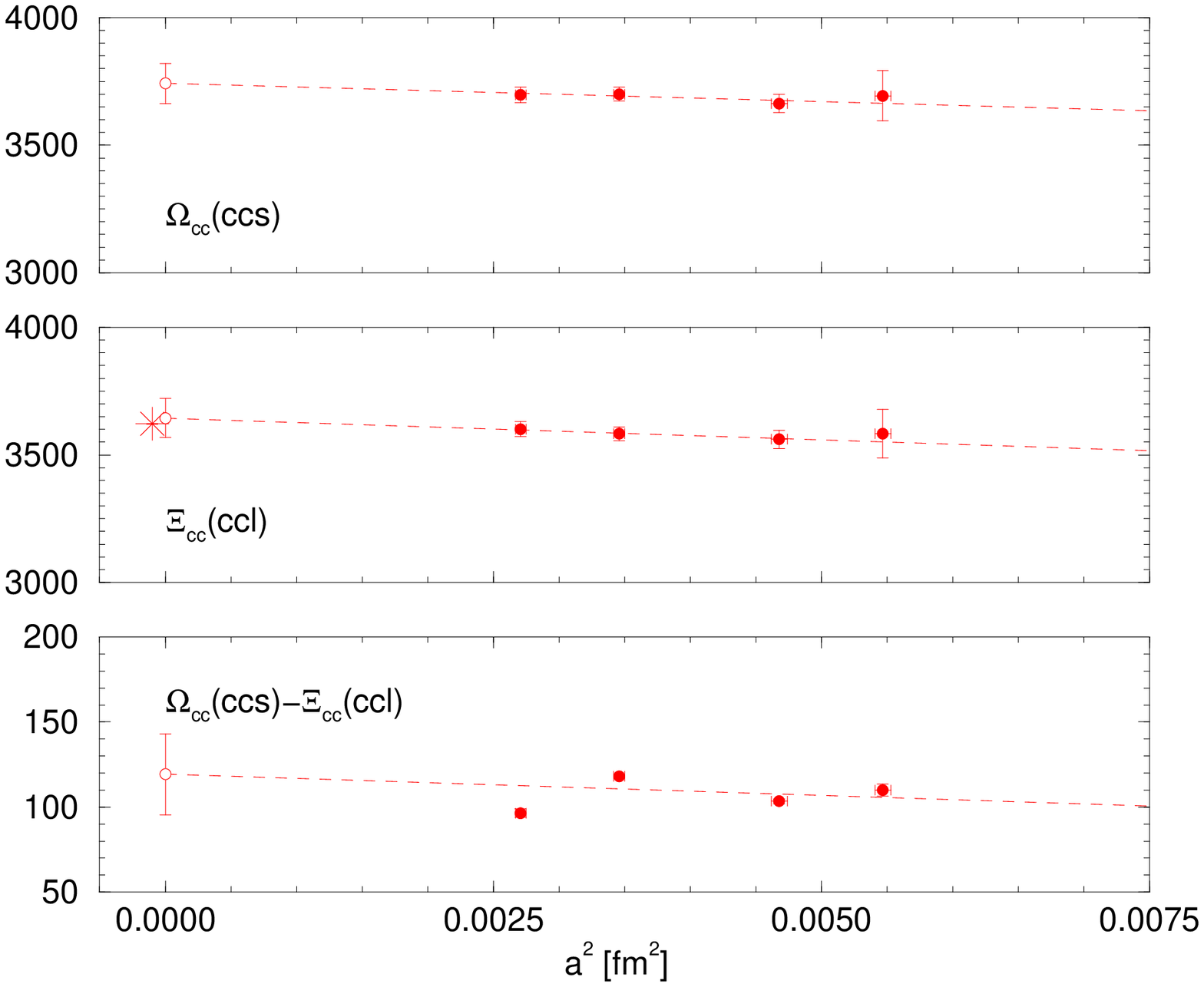}
   \end{center} 

\end{minipage}

\end{center}
\caption{Left panel: Preliminary $\Omega_c(ssc)$, $\Sigma_c(llc)$ and mass
         difference results versus $a^2$ (upper to lower plots),
         together with linear extrapolations to the continuum limit.
         Right panel: similarly for the $C=2$ open charm states 
         $\Sigma_c(llc)$, $\Omega_c(ssc)$. The LHCb result, 
         \protect\cite{LHCb17a}, is also shown in the centre plot.}
\label{charm_baryon}
\end{figure}

In the left panel of Fig.~\ref{charm_baryon} we show the $C=1$ open
charmed baryons
$\Omega_c(ssc)$, $\Sigma_c(llc)$ masses, together with their mass
splitting. For the single open charm states there is reasonable agreement
with the experimental results. Again, as for the open charm
pseudoscalar masses, the scaling violations seem to be moderate.

\begin{figure}[!h]
   \begin{center}
      \includegraphics[width=13.00cm]
         {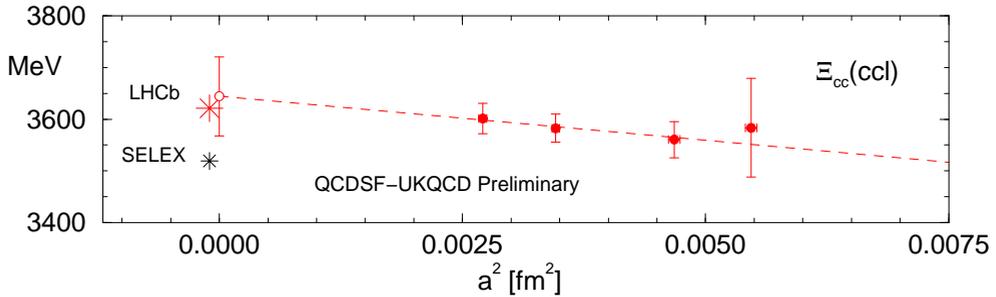}
   \end{center}
\caption{An enlarged plot for the $\Xi_{cc}$ mass from 
         Fig.~\protect\ref{charm_baryon}.
         Also indicated are the LHCb result, \protect\cite{LHCb17a}
         and the SELEX result, \cite{selex02a}.}
\label{large_scale}
\end{figure}

In the right panel of Fig.~\ref{charm_baryon} we show the $C=2$ open
charmed baryons. We examine the result for the mass of the $\Xi_{cc}$ 
in greater detail in Fig.~\ref{large_scale}. Our tentative conclusion
is that our results can differentiate between LHCb and SELEX and
tend to support the LHCb result. (As mentioned before we assume
that isospin breaking results are small.) This also is in agreement
with other recent results, such as
\cite{Namekawa:2013vu,Alexandrou:2014sha,Bali:2015lka,Alexandrou:2017xwd}.

%----------------------------------------------------------------------------

\section{Conclusions}

%----------------------------------------------------------------------------

For $u$, $d$, $s$ quarks, have developed a method to 
approach the physical point on a path starting from a point
on the SU(3) flavour symmetric line. We have developed
precise SU(3) flavour symmetry breaking expansions -- nothing is ad-hoc.
The expansions have been extended -- PQ or when mass valence quarks
$\not=$ mass sea quarks). This enables a better determination of
the expansion coefficients and also allows the expansion to be applied
in the region of the $c$ quark mass. We have data for four lattice
spacings and have applied the method to determine some open charm state
masses, in particular the recently discovered $C=2$ open charm
state $\Xi_{cc}^{++}$. The preliminary results are displayed in
Fig.~\ref{large_scale}.

In the future we plan to extend the results to the other states
shown in Table~\ref{charm_wfs} increasing the number of lattice
spacing to five. Furthermore, the expansions also are valid for
$m_u \not= m_d$, so with the determined coefficients it may be
possible to investigate pure QCD isospin breaking effects,
in particular to generalise to include mixing, \cite{horsley14a},
for example to $\Sigma_c^+$ - $\Lambda_c^+$, $\Xi_c^0$ - $\Xi_c^{\prime 0}$
mixing. Further possibilities include an investigation of the 
charmed baryon decuplet and QED effects. For the latter we are in the 
process of investigating using our recently generated ensemble of fully
dynamical QCD+QED configurations \cite{Horsley:2015eaa,Horsley:2015vla}.

%----------------------------------------------------------------------------

\section*{Acknowledgements}

%----------------------------------------------------------------------------

The numerical configuration generation (using the BQCD lattice
QCD program \cite{nakamura10a}) and data analysis 
(using the Chroma software library \cite{edwards04a}) was carried out
on the IBM BlueGene/Qs using DIRAC 2 resources (EPCC, Edinburgh, UK),
and at NIC (J\"ulich, Germany)
%, the Lomonosov at MSU (Moscow, Russia)
and the SGI ICE 8200 and Cray XC30 at HLRN (The North-German Supercomputer
Alliance) and on the NCI National Facility in Canberra, Australia
(supported by the Australian Commonwealth Government).
HP was supported by DFG Grant No. SCHI 422/10-1 and
GS was supported by DFG Grant No. SCHI 179/8-1.
PELR was supported in part by the STFC under contract ST/G00062X/1
and JMZ was supported by the Australian Research Council Grant
No. FT100100005 and DP140103067. We thank all funding agencies.

%----------------------------------------------------------------------------

%%%%%%%%%%%%%%%%%%%%%%%%%%%%%%%%%%%%%%%%%%%%%%%%%%%%%%%%%%%%%%%%%%%%%%%%%%%%%

\end{document}